\documentclass[12pt]{article}

\usepackage{graphicx}
\usepackage{cite}
\usepackage{amsmath}

\begin{document}

\title{Dynamics of ion-phosphate lattice of DNA in left-handed double helix form}
\author{S.M. Perepelytsya, S.N. Volkov\\
Bogolyubov Institute for Theoretical Physics, NAS of Ukraine,
\\14-b Metrolohichna St., Kiev, 03680, Ukraine } \maketitle

\begin{abstract}
The conformational vibrations of \emph{Z}-DNA with counterions are
studied in framework of phenomenological model developed. The
structure of left-handed double helix with counterions
neutralizing the negatively charged  phosphate groups of DNA is
considered as the ion-phosphate lattice. The frequencies and Raman
intensities for the modes of \emph{Z}-DNA with Na$^{+}$ and
Mg$^{2+}$ ions are calculated, and the low-frequency Raman spectra
are built. At the spectra range about the frequency 150 cm$^{-1}$
new mode of ion-phosphate vibrations is found, which characterizes
the vibrations of Mg$^{2+}$ counterions. The results of our
calculations show that the intensities of \emph{Z}-DNA modes are
sensitive to the concentration of magnesium counterions. The
obtained results describe well the experimental Raman spectra of
\emph{Z}-DNA.
\end{abstract}

\section{Introduction}
\label{sec:1}

DNA is macromolecule with the double helix structure which may
adopt different forms as a response to changes of environmental
conditions ~\cite{Ivanov,Saenger,Blagoi,Maleev,Williams}. Under
the natural conditions the double helix is right-handed due to the
stabilization by metal ions (counterions) that neutralize the
negatively charged phosphate groups of the macromolecule backbone.
Increasing the concentration of counterions the double helix may
take the left-handed conformation with  zigzag like backbone
(\emph{Z}-form), which is significant in many biological processes
~\cite{Rich0}. The study of counterion influence on the structure
and dynamics of the left-handed form of the double helix is
important for the understanding the mechanisms of DNA biological
functioning.

The counterions of different type may be localized in different
manner with respect to the double helix phosphate groups. The
experimental data for solid samples of DNA show that monovalent
counterions are usually localized near the oxygen atoms of the
phosphate groups from the outside of the macromolecule
~\cite{Tereshko1,Tereshko2}, while the counterions of higher
charge may neutralize the phosphate groups of different DNA
strands or bind to the nucleic bases ~\cite{Hud}. In the solution
the counterions do not have defined positions, and they form a
shell near DNA surface ~\cite{Manning,Frank,Levin,Kornyshev}. In
experiment this shell is observed as a cloud around the double
helix ~\cite{Das,Andersen1,Andersen2,Qiu}. The molecular dynamics
simulations of DNA counterions in water solution show that the
residence time of counterions near the phosphate group of DNA
backbone is about 1 ns ~\cite{Zakrzewska,Beveridge}. In this time
scale the counterions may form a regular structure around the
double helix. Such structure of counterions and DNA phosphate
groups may be considered as a lattice of ionic type (ion-phosphate
lattice).

The DNA ion-phosphate lattice is expected to have properties of
ionic crystals, therefore it should be characterized by counterion
vibrations with respect to the phosphate groups (ion-phosphate
vibrations). The ion-phosphate vibrations must be coupled with the
internal vibrations of the double helix, since the dynamics of the
ion-phosphate lattice is the part of DNA conformational dynamics.
The determination of DNA ion-phosphate modes is of paramount
importance for the understanding of counterion influence on the
structure and dynamics of DNA double helix.

The DNA ion-phosphate vibrations may be observed in the
low-frequency spectra (10$\div$200 cm$^{-1}$) where the ion
vibrations in ionic crystals and electrolyte solution are
prominent ~\cite{Kittel,Electrolyte}. In this spectra range the
modes of DNA conformational vibrations are also observed
~\cite{Powell,Lamba,Weidlich3,Weidlich4}, which are characterized
by the displacements of the atomic groups in the nucleotide pairs
(phosphates, nucleosides and nucleic bases) from their equilibrium
positions ~\cite{VK1,VK3,VK4,VK5}. The conformational vibrations
of the right-handed forms of the double helix are described well
within the framework of the phenomenological approach
~\cite{VK1,VK3,VK4,VK5}. In our previous works
~\cite{PV1,PV2,PV3,PV4,PV5,PV6} this approach has been extended
for the case of vibrations of the monovalent couterions with
respect to the phosphate groups, and the ion-phosphate vibrations
are determined for the right-handed double helix with Na$^{+}$,
K$^{+}$, Rb$^{+}$, and Cs$^{+}$ counterions. The results have been
showed that the frequencies of ion-phosphate vibrations decrease
as counterion mass increases from 180 to 90 cm$^{-1}$. The
character of DNA conformational vibrations is found to be very
sensitive to the counterion type and localization with respect to
the double helix atomic groups ~\cite{PV5,PV6}.

In case of \emph{Z}-DNA  the character of conformational
vibrations may be essentially different because the structure of
the left-handed double helix has the unique features. The
phosphate groups of the left-handed double helix form a backbone
with zigzag shape and the nearest base pairs have different
overlapping. Therefore, the nucleotide pairs form the  dimers
which are the elementary units of   \emph{Z}-DNA ~\cite{Saenger}.
The overlapping of the base pairs inside dimer is stronger than
between dimers. In the same time, the dimeric structure of
\emph{Z}-DNA is also stabilized by counterions (usually Mg$^{2+}$)
that may be localized between phosphate groups of different
strands of the double helix. The formation of dimers by nucleotide
pairs and the localization of counterions in the DNA minor grove
may be essential for the low-frequency modes of \emph{Z}-DNA. The
experimental data show that in the low-frequency spectra of
\emph{Z}-DNA new modes near 150 and 42 cm$^{-1}$ are observed
~\cite{WeidlichZ}. The origin of these modes is not determined
yet. To describe the low-frequency spectra of \emph{Z}-DNA the
approach ~\cite{PV1,PV2,PV3,PV4,PV5,PV6} should be extended
considering the features of the structure of zigzag like form of
the double helix and possible cases of counterion localization
with respect to the phosphate groups.

The goal of the present work is to find the modes of optic type
for \emph{Z}-DNA with counterions and to describe the
conformational dynamics of the left-handed double helix. To solve
this problem in section 2 the approach for the description of
dynamics of DNA ion-phosphate lattice
~\cite{PV1,PV2,PV3,PV4,PV5,PV6} is extended for the case of
\emph{Z}-DNA. In section 3 the frequencies and the Raman
intensities of the modes of DNA conformational vibrations are
calculated and the low-frequency spectrum of \emph{Z}-DNA is
built. Using the  calculated spectrum the origin of new modes of
\emph{Z}-DNA observed in the experimental spectra is determined.

\section{Model for conformational vibrations of ion-phosphate lattice of \emph{Z}-DNA}
\label{sec:2}

To describe the conformational dynamics of DNA ion-phosphate
lattice the typical cases of counterion localization with respect
to the phosphate groups should be considered. Usually the
monovalent counterions are localized from the outside of the
double helix, where one monovalent counterion neutralizes one
phosphate group (single-stranded position of counterion). In the
same time, due to the zigzag like shape of the backbone of
\emph{Z}-DNA (Fig. 1a), the distance between phosphate groups of
different strands  is rather short that makes possible the
localization of counterions in the minor grove of the left-handed
double helix between phosphates of different strands
(cross-stranded position of counterion). The cross-stranded
position is more favorable for bivalent counterions (Fig. 1a).

To describe the conformational vibrations of DNA double helix  the
phenomenological approach ~\cite{VK1,VK3,VK4} is used. In
framework of this approach the phosphate groups
(PO$_{4}$+C$_{5'}$) and nuclosides are modelled as masses $m_{0}$
and  $m$, respectively. The nucleosides rotate as the physical
pendulums with respect to the phosphate groups in plane of
nucleotide pair. The physical pendulums are characterized by
reduced length $l$. The nucleosides of different chains are paired
by H-bonds (Fig. 1b). The motions of structural elements of the
monomer link are considered in the plane orthogonal to the helical
axis (transverse vibrations). The longitudinal vibrations of the
macromolecule atomic groups have much higher frequencies and are
beyond the scope of this work.

\begin{figure}
\begin{center}
\resizebox{0.6\textwidth}{!}{%
  \includegraphics{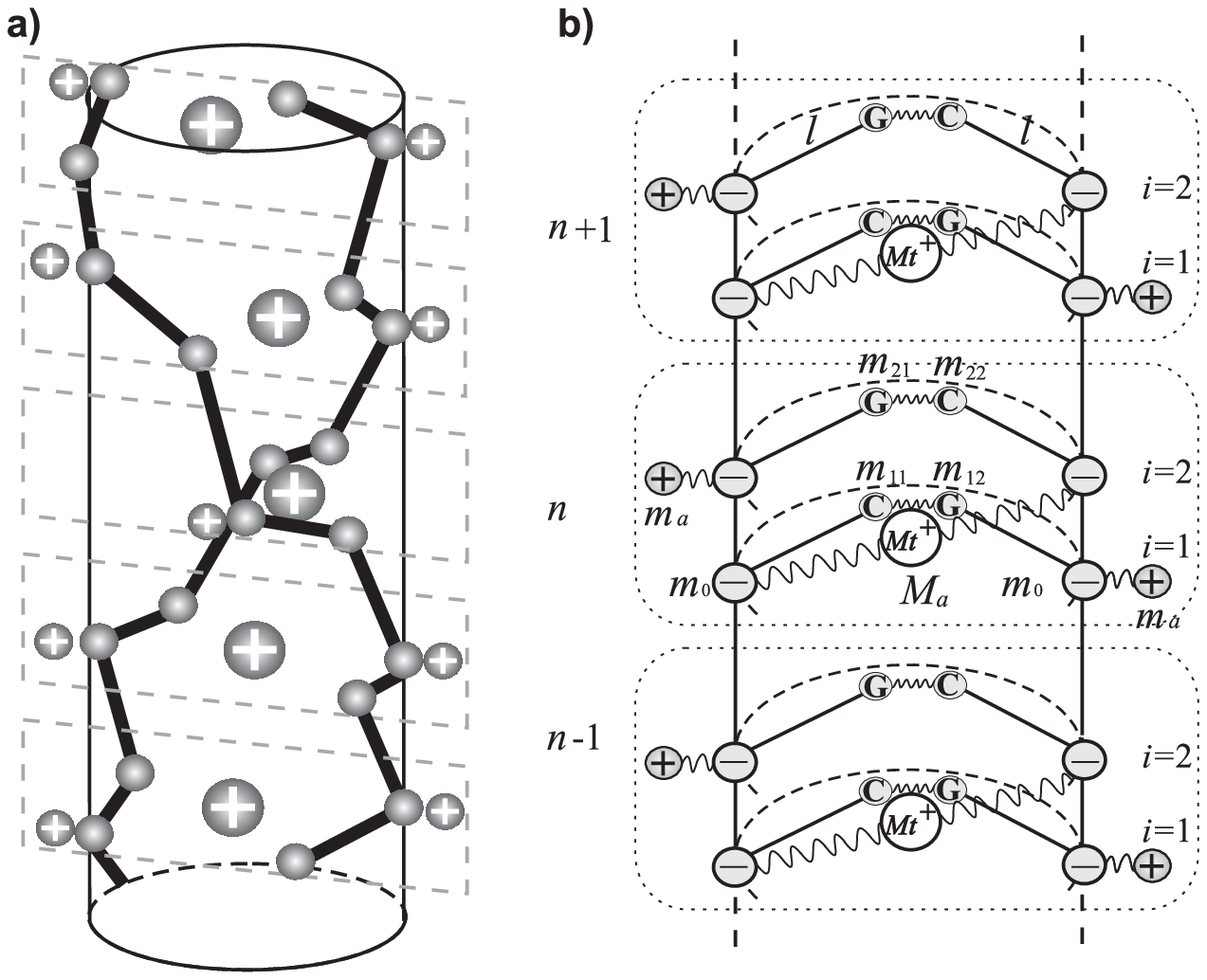}
}
\caption{Ion-phosphate lattices of \emph{Z}-DNA. (a) Zigzag-like
shape of \emph{Z}-DNA backbone. (b) Model for conformational
vibrations of the left-handed double helix with counterions. The
monomer links of the model consisted of two nucleotide pairs
(dimers) are showed by dotted frames.} \label{fig1}
\end{center}
\end{figure}

According to the dimeric structure of  the zigzag lake double
helix the monomer link of the model of \emph{Z}-DNA consists of
two nucleotide pairs. The nucleosides in dimer usually have
alternated sequence of guanine (G) and cytosine (C) nucleic bases
~\cite{Saenger}. The displacements of nucleosides and phosphate
groups in DNA monomer link are described by coordinate $Y$. The
coordinate $\theta$ describes the deviations of
pendulum-nucleosides from their equilibrium position (angle
$\theta_{0}$) in the plane of complementary DNA pair. The
vibrations of deoxyribose and base with respect to each other,
inside the nucleoside (intranucleoside mobility), are described by
changes of pendulum lengths $\rho$. The vibrations of counterion
in single-stranded positions are described by coordinate $\xi$.
For description of vibrations of a counterion in cross-stranded
positions the coordinate $Y_{a}$ is used. The vibrational
coordinates of the model and the positive directions of
displacements are showed on Figure 2.

\begin{figure}
\begin{center}
\resizebox{0.5\textwidth}{!}{%
  \includegraphics{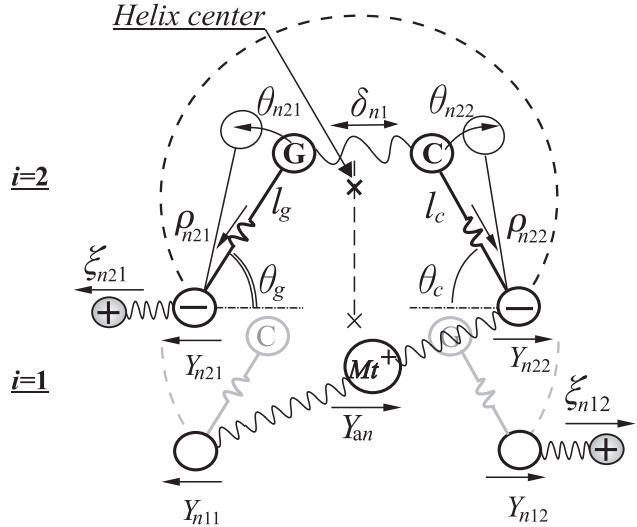}
}
\end{center}
 \caption{Monomer link of the left-handed
double helixes (dimer of nucleotide pairs).  $l$ is reduced length
of pendulum-nucleoside; $\theta_{0}$ is equilibrium angle; $m$,
$m_{0}$, and $m_{a}$ are masses of nucleosides, phosphate groups,
and counterions, respectively; $X$, $Y$, $\theta$, $\rho$, and
$\xi$ are vibrational coordinates of the model (see text). The
arrows indicate positive directions of displacements.}
\label{fig2}       
\end{figure}

Within the framework of introduced model of the dynamics of DNA
ion-phosphate lattice the energy of vibrations of structural
elements of the double helix may be written as follows:

\begin{equation}\label{Eq1}
  E=\sum_{n}\left(K_{n}+U_{n}+U_{n,n-1}\right),
\end{equation}
where $K_{n}$ and $U_{n}$ are the kinetic and potential energies
of the monomer link $n$ of DNA ion-phosphate lattice, $U_{n,n-1}$
is the potential energy of interaction along the chain. The
kinetic energy of the monomer link may be written as follows:

\begin{equation}\label{Eq2}
K_{n}=K_{0n}+K_{an},
\end{equation}
where $K_{0n}$ and $K_{an}$ are the energies of vibrations of
structure elements in nucleotide pairs of $n$-th dimer and
counterions, respectively. The potential energy of displacements
of the masses in monomer link may be written as follows:

\begin{equation}\label{Eq3}
U_{n}=U_{0n}+U_{an},
\end{equation}
where $U_{0n}$ and $U_{an}$ are the energy of vibrations of the
masses in nucleotide pairs of dimer and the energy of vibrations
counterions, respectively.

The expressions for $K_{0n}$ and $U_{0n}$ for \emph{Z}-DNA may be
written in the following form:
\begin{align}\label{Eq4}
  K_{0n}=\frac{1}{2}\sum_{i}^2\sum_{j}^2[M_{ij}\dot{Y}_{nij}^2+m_{ij}(\dot{\rho}_{nij}^2+l^2_{ij}\dot{\theta}_{nij}+
  2l_{ij}^{s}\dot{\theta}_{nij}\dot{Y}_{nij}+2b_{ij}\dot{\rho}_{nij}\dot{Y}_{nij})],
\end{align}

\begin{align}\label{Eq5}
  U_{0n}
  =\frac{1}{2}\sum_{i}^2[\alpha\delta_{ni}^{2}+\sum_{j}^{2}(\sigma_{ij}\rho_{nij}^{2}+\beta_{ij}\theta_{nij}{^2})]+
\frac{1}{2}\sum_{j}^{2}[g_{\theta}(\theta_{n1j}-\theta_{n2j})^{2}+g_{\rho}(\rho_{n1j}-\rho_{n2j})^{2}+\\\notag
+g_{y}(Y_{n1j}-Y_{n2j})^{2}],
\end{align}
where $l_{ij}^{s}=l_{ij}a_{ij}$; $a_{ij}=\sin{\theta_{0ij}}$;
$b_{ij}=\cos{\theta_{0ij}}$; $n$ enumerates dimers of
macromolecule; $i=1,2$ enumerates the nucleotide pairs in these
dimers; $j=1,2$ enumerates the chain of the double helix;
$\alpha$, $\sigma_{ij}$, and $\beta_{ij}$ are the force constants
describing H-bond stretching in base pairs, intranucleoside
mobility, and rotation of nucleosides with respect to the backbone
chain in base-pair plane, respectively; $g_{\theta}$, $g_{\rho}$,
and $g_{y}$ are the force constants describing the interaction of
nucleic bases and the interaction between phosphate groups of
different nucleotide pairs in dimers. The variable $\delta_{ni}$
describes stretching of H-bonds in the base pairs (Fig. 2). In the
present work it is determined analogically to ~\cite{VK3,VK4}:
$\delta_{ni}\approx{l_{si1}\theta_{ni1}+l_{si2}\theta_{ni2}+Y_{ni1}+Y_{ni2}+b_{i1}\rho_{ni1}+b_{i2}\rho_{ni2}}$.

The kinetic and potential energies of counterion vibrations in the
ion-phosphate lattice of \emph{Z}-DNA may be written as follows:

\begin{equation}\label{Eq7}
K_{an}=\frac{m_{a}}{2}[(\dot{Y}_{n21}+\dot{\xi}_{n21})^{2}+(\dot{Y}_{n12}+\dot{\xi}_{n12})^{2}]+\frac{M_{a}}{2}\dot{Y}_{an}^{2},
\end{equation}

\begin{align}\label{Eq8}
U_{an}=\frac{\gamma}{2}(\xi_{n21}^{2}+\xi_{n12}^{2})+
\frac{\gamma_{a}}{2}\left[(Y_{an}-Y_{n22})^{2}+(Y_{an}+Y_{n11})^{2}\right],
\end{align}
where  $\gamma$ and $\gamma_{a}$ are the force constants for the
counterions in single-stranded and cross-stranded positions,
respectively; $m_{a}$ and $M_{a}$ are the masses of counterions in
single-stranded and cross-stranded positions, respectively. In
equation (6) and (7) the first terms describe the kinetic and
potential energy of vibrations of counterions in single-stranded
positions, while the last terms describe the the kinetic and
potential energy of vibrations of counterions in cross-stranded
positions.

To describe the low-frequency Raman spectra of \emph{Z}-DNA we
intent to find the optic type phonons, which are observed in the
experimental spectra from 10 to 200 cm$^{-1}$. Therefore, we will
consider the modes characterizing the vibrations of DNA atomic
groups inside the \emph{Z}-DNA monomers (dimers of nucleotide
pairs) which are prominent in the considered spectra range. The
vibrations of the pair dimers with respect to each other by our
estimations occur with essentially lower frequencies ($<$ 5
cm$^{-1}$). Since only the long-range optic vibrations are
observed in the low-frequency Raman spectra we will consider the
limited long-wave ($\bar{k}\rightarrow$0) vibrational modes of the
ion-phosphate lattice in \emph{Z}-DNA. From the point of the
theory of the lattice vibrations such approximation is the same as
the neglecting interaction along the chain ~\cite{Kosevich}. So,
in the following consideration we neglect the interaction term:
$U_{n,n-1}\approx0$. Within the framework of this approximation
the equations of motions may be written as follows:

\begin{equation}\label{Eq8}
  \frac{d}{dt}\frac{\partial{K_{0n}}}{\partial{\dot{q}_{n}}}+\frac{d}{dt}\frac{\partial{K_{an}}}{\partial{\dot{q}_{n}}}-\frac{\partial{U_{0n}}}{\partial{q_{n}}}-\frac{\partial{U_{an}}}{\partial{q_{n}}}=0,
\end{equation}
where $q_{n}$ denotes some vibrational coordinate in the monomer
link of model (Fig. 2). The equations of motion (\ref{Eq8}) in
explicit form for the model coordinates of \emph{Z}-DNA double
helix are showed in Appendix.

To estimate the frequencies of \emph{Z}-DNA conformational
vibrations the force constants $\alpha=85$ kcal/mol\AA$^{2}$,
$\beta=40(46)$ kcal/mol, $\sigma=43(22)$ kcal/mol\AA$^{2}$ are
used the same as in \emph{B}(\emph{A})-DNA ~\cite{VK1,VK3}. The
values of $\beta$ and $\sigma$ in case of the left-handed
\emph{Z}-DNA are different for G and C nucleosides, and they are
taken the same as in \emph{A}- and \emph{B}-forms, respectively.
The constants $g_{\theta}$ and $g_{\rho}$ are equal to 20
kcal/mol, while the constant $g_{y}=8$ kcal/mol\AA$^{2}$,
~\cite{VK1,VK3}. The structure parameters of the model ($l$ and
$\theta_{0}$) for \emph{Z}-DNA are determined using the X-ray data
(pdb code: 1dcg) ~\cite{ZX}.

The constants of ion-phosphate vibrations have been determined in
our previous works ~\cite{PV2,PV6}, by using the theory of ionic
crystals. It has been showed that the constant of ion-phosphate
vibrations depends mostly on dielectric constant of DNA
ion-hydrate shell, the Madelung constant of the system describing
the electrostatic interaction of the ion with all charges of the
system,  and parameters of the ion (Pauling radius of cation and
oxygen atom). The dielectric constant has been determined
considering the properties of the hydration shell of DNA with the
counterions of different type. The Madelung constant has been
calculated from the structure of the double helix. As the result
the values of $\gamma$ and $\gamma_{a}$ constants have been
calculated for the case of ion-phosphate lattice of DNA with
alkali metal counterions in single-stranded position and for
magnesium counterions in cross-stranded position. The magnesium
counterions are considered with the hydration shell because the
size of hydrated Mg$^{2+}$ ion corresponds to the distances
between phosphate groups of the left-handed double helix. The
hydration shell of magnesium ion consists of 4 water molecules
that are strongly bond with the ion ~\cite{Ismailov}. Thus, the
constants of ion-phosphate vibrations have the following values:
$\gamma=52$ kcal/mol \AA$^{2}$ in case of Na$^{+}$ counterions
~\cite{PV2}, while in case of Mg$^{2+}$ $\gamma_{a}=62$ kcal/mol
\AA$^{2}$ ~\cite{PV6}. Using such parameters the frequencies of
\emph{Z}-DNA conformational vibrations are estimated by the
formulae (\ref{Eq1}) -- (\ref{Eq8}).

\section{The low-frequency Raman spectra of \emph{Z}-DNA}
\label{sec:5}

According to the character of vibrations of structural elements in
DNA nucleotide pairs the obtained modes may be classified as the
modes of ion-phosphate vibrations (\textbf{Ion}), H-bond
stretching modes (\textbf{H}), modes of intranucleoside vibrations
(\textbf{S}), and modes of backbone vibrations (\textbf{B}). To
characterize the influence of counterion mobility  the frequencies
of \emph{Z}-DNA conformational vibrations for the dimers without
Mg$^{2+}$ counterions are also calculated. The frequency values
for \emph{Z}-DNA conformational vibrations are showed in the Table
1.

The calculations show that the frequencies of ion-phosphate
vibrations depend on counterion position. In case of Na$^{+}$
counterions in single-stranded position there are two degenerated
modes ($\omega_{Ion1}$ and $\omega_{Ion2}$) with the frequencies
at the top of DNA low-frequency spectra (near 180 cm$^{-1}$).
These modes are characteristic for the both right- and left-handed
double helix forms. The vibrations of hydrated Mg$^{2+}$ ions in
cross-stranded position ($\omega_{Ion3}$)  have essentially lower
frequency that is due to the big size and big mass  of hydrated
Mg$^{2+}$ ion ($M_{a}=96$ a.u.m.). The vibrations of Mg$^{2+}$
counterion in cross-stranded position do not influences the
vibrations of Na$^{+}$ counterions  outside of the double helix.

The  H-bond stretching in nucleotide pairs of \emph{Z}-DNA  are
characterized by 4 modes ($\omega_{H1}$, $\omega_{H2}$,
$\omega_{H3}$, and $\omega_{H4}$), while  in the case of the
right-handed form of the double helix there are only two modes.
Additional two modes of H-bond stretching $\omega_{H2}$ and
$\omega_{H3}$ appear due to the formation of dimers of nucleotide
pairs in \emph{Z}-DNA structure. The modes $\omega_{H1}$ and
$\omega_{H2}$ have practically the same frequency for \emph{B}-
and \emph{Z}-DNA. The modes $\omega_{H3}$ and $\omega_{H4}$ of
\emph{Z}-DNA are sensitive to Mg$^{2+}$ counterion in
cross-stranded position.

\begin{table}
\caption{Frequencies of \emph{Z}-DNA conformational vibrations
(cm$^{-1}$). The results for two cases are showed: \emph{Z}-DNA
with Na$^{+}$ and Mg$^{2+}$ counterions, and \emph{Z}-DNA with
Na$^{+}$ counterions only. The calculated frequencies are compared
with the experimental data for \emph{Z}-DNA
~\cite{Lamba,WeidlichZ} and with the calculation data for
\emph{B}-DNA ~\cite{PV2}. }
\begin{center}
\begin{tabular}{ll|cc|cc|c}\hline\noalign{\smallskip}
&&\multicolumn{4}{c|}{\emph{Z}-DNA} & \emph{B}-DNA\\\hline

&&
\multicolumn{2}{c|}{Theory}&\multicolumn{2}{c|}{Experiment$^{*}$}& Theory ~\cite{PV2}\\
\textbf{Mode}&&(Na$^{+}$,
Mg$^{2+}$)&(Na$^{+}$)&~\cite{WeidlichZ}&~\cite{Lamba}&(Na$^{+}$)\\\hline

&$\omega_{Ion1}$&180&180&--&--&181\\
\textbf{Ion}&$\omega_{Ion2}$&180&180&--&--& 181\\
&$\omega_{Ion3}$& 152&--&153w&--&--\\\hline

&$\omega_{H1}$& 110&107& 112s&116sh &110\\
\textbf{H}&$\omega_{H2}$ &  106&106&--&100& -- \\
&$\omega_{H3}$&96&68&--&--& -- \\
&$\omega_{H4}$&59&66&70&66&58\\\hline

&$\omega_{S1}$&  45&45&45vw&--& 79 \\
\textbf{S}&$\omega_{S2}$ & 44&41&--&--& -- \\
&$\omega_{S3}$&40&28&--&--& -- \\\hline

&$\omega_{B1}$& 25&25&25&--&16\\
\textbf{B}&$\omega_{B2}$ & 20&20&--&--&14 \\
&$\omega_{B3}$& 11&11&--&--&--\\
&$\omega_{B4}$&11&11&--&--&-- \\\hline
\end{tabular}
\end{center}
$^{*}$ s -- strong, sh -- shoulder, w -- weak, vw -- very weak.
\end{table}

\begin{figure}
\begin{center}
\resizebox{0.5\textwidth}{!}{%
  \includegraphics{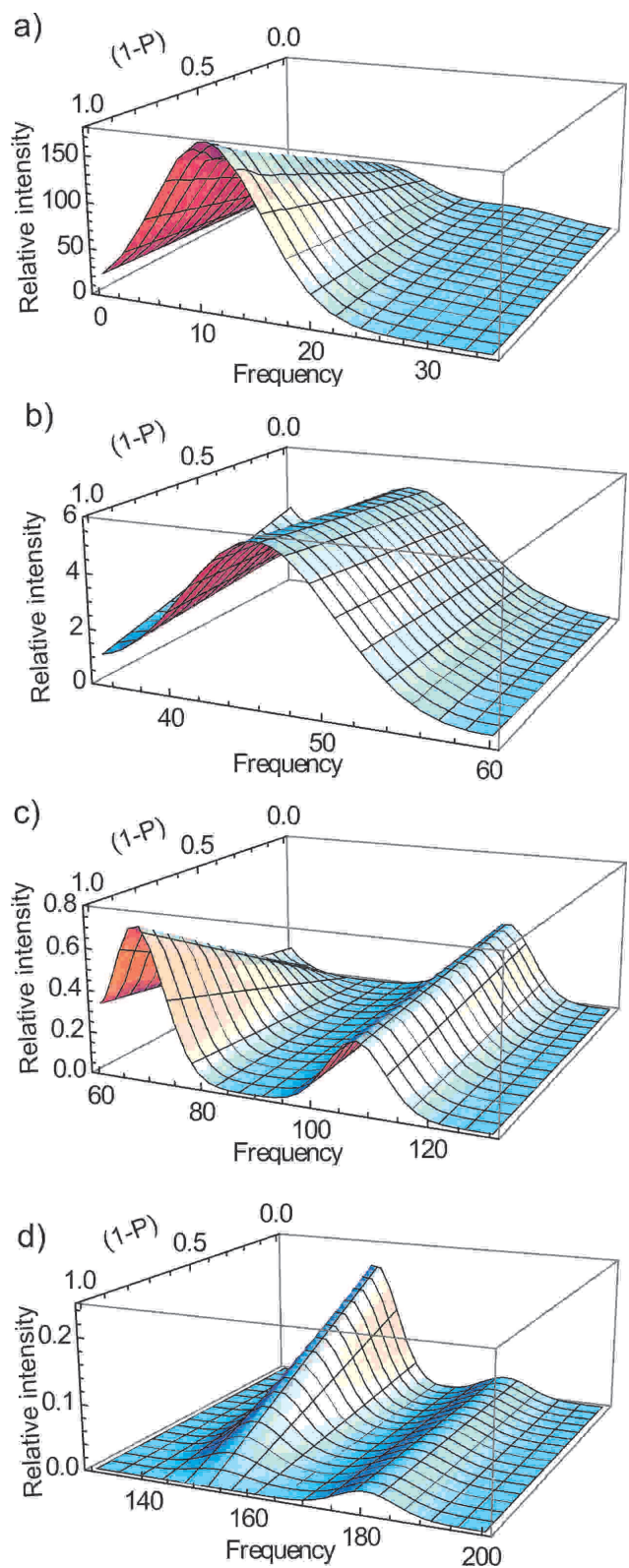}
}
\caption{The low-frequency Raman spectra of \emph{Z}-DNA with
different filling of the ion-phosphate lattice by Mg$^{2+}$
counterions. a)  Backbone vibration range (0$\div$40)
cm$^{-1}$(\textbf{B}). b) Intranucleoside vibrations range
(40$\div$60) cm$^{-1}$ (\textbf{S}). c) H-bond stretching
vibrations range (60$\div$130) cm$^{-1}$ (\textbf{H}). d)
Ion-phosphate vibrations range (130$\div$200) cm$^{-1}$
(\textbf{Ion}).} \end{center}
\label{fig2}       
\end{figure}

In case of left-handed double helix there are 3 modes of
intranucleoside vibrations $\omega_{S1}$, $\omega_{S2}$, and
$\omega_{S3}$ with rather close frequency values (about 45
cm$^{-1}$). The modes of intranucleoside vibrations depend
essentially  on the double helix form. In case of  \emph{B}-DNA
there is only one mode of intranucleoside vibrations
$\omega_{S1}$, which has essentially higher frequency than
respective modes of \emph{Z}-DNA.  The mode $\omega_{S3}$ shifts
to lower frequencies if there is no Mg$^{2+}$ counterion in
cross-stranded position.

The modes of backbone vibrations ($\omega_{B1}$, $\omega_{B2}$,
$\omega_{B3}$, and $\omega_{B4}$) are the lowest in DNA
low-frequency spectra. In case of the ion-phosphate lattice of
\emph{Z}-DNA there are two modes ($\omega_{B1}$, $\omega_{B2}$)
near 25 cm$^{-1}$ and two modes ($\omega_{B3}$, $\omega_{B4}$)
with close frequency values (about 10 cm$^{-1}$). In case of the
ion-phosphate lattice of \emph{B}-DNA there are only two modes
$\omega_{B1}$ and $\omega_{B2}$ with close frequency values. This
difference of \emph{B}- and \emph{Z}-DNA spectra is the
manifestation of the dimeric structure of the left-handed
structure of  DNA double helix.

To compare  our calculations with the experimental spectra the
Raman intensities are calculated using the approach developed in
~\cite{PV5,PV6}. As the result the low-frequency Raman spectra of
DNA in \emph{Z}-form of the double helix are built for different
degree of filling ($P$) of the ion-phosphate lattice by Mg$^{2+}$
counterions (Fig. 3). The parameter $P$ equals to 1 if all dimers
of \emph{Z}-DNA contain Mg$^{2+}$ counterions, and $P$ equals to 0
if there is no counterions in dimers. The mode intensities are
normalized per intensity of the mode $\omega_{H1}$. The halfwidth
of spectra lines is considered equaled to 5 cm$^{-1}$

The obtained spectrum is divided into four ranges with similar
intensities of the modes of DNA conformational vibrations (Fig. 3a
-- 3d). The range (0$\div$40) cm$^{-1}$ (Fig. 3a) is characterized
by the modes of backbone vibrations (\textbf{B}). This modes are
the most intensive in DNA low-frequency spectra because their
energy is lower than $k_{B}T$ at room temperatures. Due to the
high intensity the modes of backbone vibrations are observed in
the experimental spectra as a peak near 25 cm$^{-1}$
~\cite{WeidlichZ}. Increasing the filling of the ion-phosphate
lattice   by Mg$^{2+}$ counterions in the dimers the intensity of
this band decreases.

At frequency range (40$\div$60)  cm$^{-1}$ (Fig. 3b) there is a
band characterizing the intranucleoside vibrations (\textbf{S}).
The intensity of this band is essentially lower than the intensity
of \textbf{B} band formed by modes of backbone vibrations.
Therefore in the experimental spectra ~\cite{WeidlichZ} \textbf{S}
band is observed as a weak mode near 45 cm$^{-1}$. The dependence
of intensity of this band on filling parameter $P$ is weak.

The frequency range (60$\div$130) cm$^{-1}$  (Fig. 3c) is
characterized by two bands 70 and 110 cm$^{-1}$ that are caused by
the modes of H-bond stretching in the base pairs (\textbf{H}). The
intensity of the band 70 cm$^{-1}$ decreases as the filling of the
ion-phosphate lattice by Mg$^{2+}$ counterions increases, while
the intensity of the band 110 cm$^{-1}$ remain the same. The
comparison of the calculated spectra (Fig. 3c) with the
experimental Raman spectra ~\cite{Lamba,WeidlichZ} show that at
these frequencies there are two strong modes. The frequency and
intensity of the observed modes agree with our calculations.

At the spectra range (130$\div$200) cm$^{-1}$ (Fig. 3d) there are
two bands 150 and 180 cm$^{-1}$ characterizing the ion-phosphate
vibrations of Mg$^{2+}$ and Na$^{+}$ counterions, respectively
(\textbf{Ion}). The dependence of intensity on filling parameter
$P$ is essential only in case of the mode of ion-phosphate
vibrations of Mg$^{2+}$ counterions. The comparison with the
experimental spectra show that in the Raman spectra of
\emph{Z}-DNA very weak mode near 153 cm$^{-1}$ is observed
~\cite{WeidlichZ} that corresponds to the calculated band 150
cm$^{-1}$. The vibrations of Na$^{+}$ counterions are not observed
in the low-frequency spectra of \emph{Z}-DNA. However, in the
spectra of dry DNA films the mode resembling the mode of
ion-phosphate vibrations has been detected
~\cite{Powell,Weidlich4,PV1,PV2}.

\section{Conclusions} \label{sec7}

The conformational vibrations of \emph{Z}-DNA   are studied
considering the double helixes with counterions as the lattice of
ionic type (ion-phosphate lattice). To find vibrational modes of
the ion-phosphate lattice the phenomenological model is developed
taking into consideration the feature of \emph{Z}-DNA  structure.
In the model the monovalent counterions are localized outside of
the double helix, and bivalent counterions are localized between
phosphate groups of nucleotide dimers of \emph{Z}-DNA. Using the
developed model the frequencies and the Raman intensities of
vibrational modes are calculated for DNA with Na$^{+}$ and
Mg$^{2+}$ counterions. The results show that the obtained
frequencies of H-bond stretching in nucleotide pairs and the
frequencies of intranucleoside vibrations differ from respective
frequencies in case  of the right-handed DNA double helix. The
reason of such difference  is in the dimeric structure of the
left-handed \emph{Z}-DNA. New mode of ion-phosphate vibrations
about the frequency 150 cm$^{-1}$ is determined, which
characterizes the vibrations of Mg$^{2+}$ counterions. The
intensity of this mode is rather large, and it is observed in the
experimental Raman spectra of \emph{Z}-DNA ~\cite{WeidlichZ}. The
modes of ion-phosphate vibrations in case of Na$^{+}$ counterions
(180 cm$^{-1}$) are weak and do not observed experimentally. The
modes of \emph{Z}-DNA conformational vibrations are very sensitive
to the concentration of Mg$^{2+}$ counterions in the solution. The
developed model describes the conformational vibrations of
\emph{Z}-DNA with Na$^{+}$ and Mg$^{2+}$ counterions and allows
determine the origin of new modes in the experimental
low-frequency Raman spectra of \emph{Z}-DNA.

\begin{flushleft}
{\footnotesize The present work was partially supported by the
Grant for the Young Scientists of the NAS of Ukraine 0112U005857
and by the project  of the NAS of Ukraine 0110U007540}
\end{flushleft}

\section*{Appendix: Equations of motion}

For a more convenient form of the equations of motion the
following variables   are used: $ Y_{c}^{n}=Y_{11}^{n}+Y_{22}^{n},
\quad y_{c}^{n}=Y_{11}^{n}-Y_{22}^{n}, \quad
\theta_{c}^{n}=\theta_{11}^{n}+\theta_{22}^{n}, \quad
\eta_{c}^{n}=\theta_{11}^{n}-\theta_{22}^{n}, \quad
\rho_{c}^{n}=\rho_{11}^{n}+\rho_{22}^{n}, \quad
r_{c}^{n}=\rho_{11}^{n}-\rho_{22}^{n}, \quad
Y_{g}^{n}=Y_{21}^{n}+Y_{12}^{n}, \quad
y_{g}^{n}=Y_{21}^{n}-Y_{12}^{n}, \quad
\theta_{g}^{n}=\theta_{21}^{n}+\theta_{12}^{n}, \quad
\eta_{g}^{n}=\theta_{21}^{n}-\theta_{12}^{n}, \quad
\rho_{g}^{n}=\rho_{21}^{n}+\rho_{12}^{n}, \quad
r_{g}^{n}=\rho_{21}^{n}-\rho_{12}^{n}, \quad
\xi_{1}^{n}=\xi_{21}^{n}+\xi_{12}^{n}, \quad
\xi_{2}^{n}=\xi_{21}^{n}-\xi_{12}^{n}.$ As the result the
equations of motion (\ref{Eq8}) for the optic long wave vibrations
are split into two subsystems of coupled equations:

\[\left\{\begin{array}{l}

{\ddot{Y}_{c}^{n} +\frac{m_{c} b_{c} }{M_{c} } \rho _{c}^{n}
+\frac{m_{c} l_{c}^{s} }{M_{c} } \ddot{\theta }_{c}^{n} +
\frac{M_{a} }{M_{c} }
\gamma _{a0} Y_{c}^{n} -g_{yc} (Y_{c}^{n} -Y_{g}^{n})+}\\
{\alpha _{c} \left(Y_{g}^{n} +Y_{c}^{n} +l_{c}^{s} \theta _{g}^{n} +l_{c}^{s} \theta _{c}^{n} +b_{g} \rho _{g}^{n} +b_{c} \rho _{c}^{n} \right)=0\, ;} \\

{\ddot{Y}_{g}^{n} +\frac{m_{g} b_{g} }{M_{g} } \rho _{g}^{n}
+\frac{m_{g} l_{g}^{s} }{M_{g} } \ddot{\theta }_{g}^{n}
+\frac{m_{a} }{M_{g} }
(\ddot{Y}_{g}^{n} +\ddot{\xi}_{1}^{n} )-g_{yg} (Y_{c} ^{n}-Y_{g}^{n})+}\\
{\alpha _{g} \left(Y_{g}^{n} +Y_{c}^{n} +l_{g}^{s} \theta _{g}^{n}
+l_{c}^{s}
\theta_{c}^{n} +b_{g} \rho _{g}^{n} +b_{c} \rho _{c}^{n} \right)=0\, ;}\\

{\ddot{\theta }_{g}^{n} +\ddot{Y}_{g}^{n} \frac{a_{g} }{l_{g} }
+\beta _{g} \theta _{g}^{n} -g_{\theta g} (\theta _{c}^{n} -\theta
_{g}^{n})+\alpha _{g} \frac{M_{g} a_{g} }{m_{g} l_{g} }
\left(Y_{g}^{n} +Y_{c}^{n} +l_{c}^{s} \theta _{g}^{n} +l_{c}^{s}
\theta _{c}^{n} +b_{g} \rho _{g}^{n} +b_{c} \rho _{c}^{n}
\right)=0\, ;}

\\ {\ddot{\theta }_{c}^{n} +\ddot{Y}_{c}^{n}
\frac{a_{c} }{l_{c} } +\beta _{c} \theta _{c}^{n} -g_{\theta
c}(\theta _{c}^{n} -\theta _{g}^{n})+\alpha _{c} \frac{M_{c} a_{c}
}{m_{c} l_{c} } \left(Y_{g}^{n} +Y_{c}^{n} +l_{g}^{s} \theta
_{g}^{n} +l_{c}^{s} \theta _{c}^{n} +b_{g} \rho _{g}^{n} +b_{c}
\rho _{c}^{n} \right)=0\, ;}

\\
{\ddot{\rho }_{g}^{n} +b_{g} \ddot{Y}_{g}^{n} +\sigma _{g} \rho
_{g}^{n} -g_{\varrho g} \left(\rho _{c}^{n} -\rho _{g}^{n}
\right)+\alpha _{g} \frac{M_{g} b_{g} }{m_{g} } \left(Y_{g}^{n}
+Y_{c}^{n} +l_{g}^{s} \theta _{g}^{n} +l_{c}^{s} \theta _{c}^{n}
+b_{g} \rho _{g}^{n} +b_{c} \rho _{c}^{n} \right)=0\, ;}

\\
{\ddot{\rho }_{c}^{n} +b_{c} \ddot{Y}_{c}^{n} +\sigma _{c} \rho
_{c}^{n} -g_{\rho c}(\rho _{c}^{n} -\rho _{g}^{n})+\alpha _{c}
\frac{M_{c} b_{c} }{m_{c} } \left(Y_{g}^{n} +Y_{c}^{n} +l_{g}^{s}
\theta _{g}^{n} +l_{c}^{s} \theta _{c}^{n} +b_{g} \rho _{g}^{n}
+b_{c} \rho _{c}^{n} \right)=0\, ;}

\\
{\ddot{Y}_{g}^{n} +\ddot{\xi}_{1}^{n} +\gamma _{0} \xi_{1}^{n}
=0\, ;}
\end{array}\right. \]

\[\left\{\begin{array}{l}

{\ddot{y}_{c}^{n} +\frac{m_{c} b_{c} }{M_{c} } \ddot{r}_{c}^{n}
+\frac{m_{c} l_{c}^{s} }{M_{c} } \ddot{\eta }_{c}^{n} -\frac{M_{a} }{M_{c} } \gamma _{a0} \left(2y_{a}^{n} -y_{c}^{n} \right)+g_{yc} \left(y_{c}^{n} +y_{g}^{n} \right)+}\\
{\alpha _{c} \left(y_{g}^{n} +y_{c}^{n} +l_{g}^{s} \eta _{g}^{n} +l_{c}^{s} \eta _{c}^{n} +b_{g} r_{g}^{n} +b_{c} r_{c}^{n} \right)=0\, ;} \\

{\ddot{y}_{g}^{n} +\frac{m_{g} b_{g} }{M_{g} } \ddot{r}_{g}^{n}
+\frac{m_{g} l_{g}^{s} }{M_{g} } \ddot{\eta }_{g}^{n} +\frac{m_{a}
}{M_{g} } (y_{g}^{n} + \xi_{2}^{n})+g_{yg} (y_{c}^{n}
+y_{g}^{n})+}\\
{\alpha _{g} \left(y_{g}^{n} +y_{c}^{n} +l_{g}^{s} \eta _{g}^{n}
+l_{c}^{s} \eta _{c}^{n} +b_{g} r_{g}^{n} +b_{c} r_{c}^{n}
\right)=0\, ;}

\\ {\ddot{\eta }_{g}^{n} +\ddot{y}_{g}^{n} \frac{a_{g} }{l_{g} } +\beta _{g} \eta _{g}^{n} +g_{\theta g} \left(\eta _{c}^{n} +\eta _{g}^{n} \right)+\alpha _{g} \frac{M_{g} a_{g} }{m_{g} l_{g} } \left(y_{g}^{n} +y_{c}^{n} +l_{g}^{s} \eta _{g}^{n} +l_{c}^{s} \eta _{c}^{n} +b_{g} r_{g}^{n} +b_{c} r_{c}^{n} \right)=0\, ;} \\

{\ddot{\eta }_{c}^{n} +\ddot{y}_{c}^{n} \frac{a_{c} }{l_{c} }
+\beta _{c} \eta _{c}^{n}+g_{\theta c} \left(\eta _{c}^{n} +\eta
_{g}^{n} \right) +\alpha _{c} \frac{M_{c} a_{c} }{m_{c} l_{c} }
\left(y_{g}^{n} +y_{c}^{n} +l_{g}^{s} \eta _{g}^{n} +l_{c}^{s}
\eta _{c}^{n} +b_{g} r_{g}^{n} +b_{c} r_{c}^{n}
\right)=0\, ;} \\

{\ddot{r}_{g}^{n} +b_{g} \ddot{y}_{g}^{n} +\sigma _{g} r_{g}^{n}
+g_{\varrho g} \left(r_{c}^{n} +r_{g} ^{n}\right)+\alpha _{g}
\frac{M_{g} b_{g} }{m_{g} } \left(y_{g}^{n} +y_{c}^{n} +l_{g}^{s}
\eta _{g}^{n} +l_{c}^{s} \eta _{c}^{n}
+b_{g} r_{g}^{n} +b_{c} r_{c}^{n} \right)=0\, ;} \\

{\ddot{r}_{c}^{n} +b_{c} \ddot{y}_{c}^{n} +\sigma _{c} r_{c}^{n}
+g_{\rho c} \left(r_{c}^{n} +r_{g}^{n} \right)+\alpha _{c}
\frac{M_{c} b_{c} }{m_{c} } \left(y_{g}^{n} +y_{c}^{n} +l_{g}^{s}
\eta _{g}^{n} +l_{c}^{s} \eta _{c}^{n}
+b_{g} r_{g}^{n} +b_{c} r_{c}^{n} \right)=0\, ;} \\

{\ddot{y}_{g}^{n} +\ddot{\xi}_{2}^{n}+\gamma _{0}\xi_{2}^{n}=0\, ;} \\

{2\ddot{y}_{a}^{n} +2\gamma _{a0} \left(2Y_{a}^{n} -y_{c}^{n}
\right)=0\, ,}
\end{array}\right. \]
where $\alpha _{g} =\alpha /M_{g} $, $\alpha _{c} =\alpha /M_{c}$,
$\beta _{g} =\beta /m_{g} l_{g}^{2}$, $\beta _{c} =\beta /m_{c}
l_{c}^{2}$, $\sigma _{g} =\sigma /m_{g}$, $\sigma _{c} =\sigma
/m_{c}$, $\gamma _{0} =\gamma /m_{a}$, $\gamma _{a0} =\gamma _{a}
/M_{a}$, $g_{\theta g} =g_{\theta} /m_{g} l_{g}^{2}$, $g_{\theta
c} =g_{\theta} /m_{c} l_{c}^{2}$, $g_{\rho g} =g_{\rho} /m_{g}$,
$g_{\rho c} =g_{\rho} /m_{c}$, $g_{yg} =g_{y} /m_{g}$, $g_{yc}
=g_{y} /m_{c}$, $a_{c}=\sin\theta_{0c}$, $a_{g}=\sin\theta_{0g}$,
$b_{c}=\cos\theta_{0c}$, $b_{g}=\cos\theta_{0g}$,
$l_{c}^{s}=l_{c}\sin\theta_{0c}$,
$l_{g}^{s}=l_{g}\sin\theta_{0g}$.

The  first subsystem of equations of motion describes symmetric
vibrations of nucleotides in the dimers, while the second
subsystem describes antisymmetric vibrations of nucleotides. The
obtained equations of motions are solved using the substitution:
$q_{n}=\tilde{q}_{n}\exp(i\omega{t})$, where $\tilde{q}_{n}$ and
$\omega$ are amplitude and frequency for some coordinate of
vibrations, respectively. As the result the equation for
frequencies of long-wave vibrations of \emph{Z}-DNA are obtained.
From the first subsystem of equations of motion the equation for
frequencies $\omega_{H1}$, $\omega_{H3}$, $\omega_{S2}$,
$\omega_{S3}$, $\omega_{B2}$, $\omega_{B4}$, and $\omega_{Ion1}$
is derived. The second subsystem of equations of motion gives the
equation for frequencies  $\omega_{H2}$, $\omega_{H4}$,
$\omega_{S1}$, $\omega_{B1}$, $\omega_{B3}$, $\omega_{Ion2}$, and
$\omega_{Ion3}$. Using the parameters discussed in the section 2
the frequency values of the modes of \emph{Z}-DNA conformational
vibrations are calculated numerically. The values of the
calculated frequencies are showed in the Table 1.

\end{document}